\documentclass{amsart}[10pt]
\usepackage{amssymb,amsmath}
\newtheorem{theorem}{Theorem}
\newtheorem*{theorem*}{theorem}
\newtheorem{lemma}{Lemma}
\newtheorem{definition}{Definition}
\renewcommand{\to}[1][]{\xrightarrow{#1}}
\newcommand{\ep}{\varepsilon}
\begin{document}

\begin{titlepage}
%\rightline{preprint numbers}
\bigskip
\begin{center}
{{\large\bf
The number of master integrals is finite} \\
% Finiteness of number of master integrals
\vglue 5pt
\vglue 1.0cm

\vspace{2mm}
{\large  A.V. Smirnov}\footnote{E-mail: asmirnov80@gmail.com}\\
{\normalsize Scientific Research Computing Center, Moscow State
University, \\ 119992 Moscow, Russia
   }\\

{\large   A.V. Petukhov}\footnote{E-mail: a.petukhov@jacobs-university.de}\\
{\normalsize
 Department of Higher Algebra of Moscow State University, 119992 Moscow, Russia\\ Jacobs University, 28759 Bremen, Germany
}
\vspace{20mm}
\vglue 0.8cm
{Abstract}}
\end{center}
\vglue 0.3cm
{\rightskip=3pc
 \leftskip=3pc
\noindent
For a fixed Feynman graph one can consider Feynman integrals with all possible powers of propagators and try to
reduce them, by linear relations, to a finite subset of integrals, the so-called master integrals. Up to now, there are
numerous examples of reduction procedures resulting in a finite number of master integrals
for various families of Feynman integrals. However, up to now it was just an empirical fact that the reduction procedure
results in a finite number of irreducible integrals.
It this paper we prove that the number of master integrals is always finite.
\vglue 0.8cm}
\end{titlepage}

\section{Introduction}
Feynman integrals over loop momenta are building blocks of  quantum-theoretical
amplitudes in the framework of perturbation theory.
After a tensor reduction based on some projectors a
given Feynman graph generates various scalar Feynman integrals
that have the same structure of the integrand with various
distributions of powers of propagators which we shall also call
{\em indices}:

\begin{eqnarray}
  \mathrm F(a_1,\ldots,a_n) &=&
  \int \cdots \int \frac{\mbox{d}^d k_1\ldots \mbox{d}^d k_h}{\mathrm E_1^{a_1}\ldots \mathrm E_n^{a_n}}
  \label{eqbn}
\end{eqnarray}
Here $k_i$, $i=1,\ldots,h$, are loop momenta
and the denominators $\mathrm E_r$ are linear combinations of constants and quadratic forms of
loop momenta $p_i=k_i, \; i=1,\ldots,h$ and
independent external momenta $p_{h+1}=q_1,\ldots,p_{h+N}=q_N$ of the graph.
This definition covers classical denominators of the form $p^2-m^2$, where $p$ is a linear
combination of internal and external momenta and $p^2=p_0^2-\vec{p}\;^2=p_0^2-p_1^2-p_2^2-p_3^2$, as well as effective field theories
and asymptotic expansions of Feynman integrals in various limits (if we
encounter propagators of the form $k v$ then we treat $v$ the same way as external momenta).

Irreducible polynomials in the numerator can be represented as
denominators raised to negative powers.
Usual prescriptions $k^2=k^2+i 0$, etc. are implied.
The dimensional regularization \cite{dimreg} with $d=4-2\ep$  is assumed.
The Feynman integrals are functions of dimension $d$, masses,
and kinematic invariants, $q_i\cdot q_j=q_{i0} q_{j0}-\vec{q}_i \vec{q}_j$ (Lorentz scalar products). If we do not have enough irreducible
denominators to represent any quadratic form of $p_i$ as a linear combination of
masses, kinematic invariants and denominators, then we are adding extra irreducible
numerators that are treated in the absolutely same way as propagators.

At the modern level of perturbative calculations, when one needs
to evaluate millions of Feynman integrals
(\ref{eqbn}), a well-known optimal strategy here is to derive, without calculation, and then
apply some relations between the given family of Feynman integrals
as recurrence relations
A well-known standard way to obtain such relations is provided
by the method of integration by parts
(IBP) \cite{IBP}.

Practically, one starts from IBP relations
\begin{eqnarray}
\int\ldots\int \mbox{d}^d k_1 \mbox{d}^d k_2\ldots
\frac{\partial}{\partial k_i}\cdot\left( p_j
\frac{1}{\mathrm E_1^{a_1}\ldots \mathrm E_n^{a_n}}
\right)   =0  \;.
\label{RR}
\end{eqnarray}
After the differentiation, resulting scalar products,
$k_i\cdot k_j$ and $k_i\cdot q_j$
are expressed in terms of the factors in the denominator,
and one arrives at IBP relations which can be written as
\begin{equation}
\sum c_i \mathrm F(a_1+b_{i,1},\ldots,a_n+b_{i,n})
=0\,,
\label{IBP}
\end{equation}
where $b_{i,j}$ are integer,
$c_i$ are polynomials in $a_j$,
$d$, masses $m_i$ and kinematic invariants,
%and the arguments of
and $\mathrm F(a_1,\ldots,a_n)$ are Feynman integrals (\ref{eqbn}) of the
given family.

One tries to use IBP relations in order to express a general dimensionally
regularized integral of the given family as a linear combination of some
`irreducible' integrals which are also called {\it master} integrals
\footnote{See chapter 5 of \cite{Smirnov:2006ry} for a review of the method of IBPs.}.

The goal of this paper is to prove the following theorem:

\begin{theorem}
\label{Phys}
The number of master integrals is always finite.
\end{theorem}

\section{Definition of master integrals}

Normally one does not use a strict definition of master integrals - in practice one simply tries to reduce Feynman integrals as much as possible - and the integrals that could not be reduced further are declared to be master integrals
\footnote{A general algorithm to reduce integrals to master integrals was initially suggested by Laporta, Gehrmann and Remiddi \cite{LGR1,LGR2}; currently
there are multiple implementations of this algorithm including three public versions (AIR by Anastasiou and Lazopoulos \cite{AIR}, FIRE by A.~Smirnov \cite{FIRE} and Reduze by Studerus \cite{Reduze}).}.
However, after fixing a set of relations and an ordering one can easily define master integrals --- this definition was introduced in \cite{masters}. Let us repeat this definition.

Feynman integrals (\ref{eqbn}) can be considered as elements of
the field of functions $\mathcal F$ of $n$ integer arguments $a_1,a_2,\ldots,a_n$
(in the case of propagator powers depending on $d$ we will shift indices and still consider $a_i$
to be integers).

After having fixed a set of relations (here we will simply take
all possible IBPs and substitute all possible values of indices) we can generate by them an
infinite-dimensional vector subspace ${\mathcal R} \subset {\mathcal F^*}$.
Now one considers the set of solutions of all those relations, that is
the intersection of the kernels of all functionals
$r \in {\mathcal R}$. This is a vector subspace of $\mathcal F$
and will be denoted by $\mathcal S$. A Feynman integral considered
as a function of the integer variables $a_1,\ldots,a_n$ is an element of the
space $\mathcal S$, for it satisfies the IBP relations and other
relations mentioned above. Formally,
${\mathcal S} =\{ f\in {\mathcal F}: \langle r,f\rangle =0
\; \forall\; r\in {\mathcal R}\}$.

After fixing an ordering on Feynman integrals, we can define what a master integral is.
(In this paper the choice of the ordering is unsignificant, for details see \cite{masters}.)
A master integral is such an integral F$(a_1,\ldots,a_n)$ that there is no element $r\in {\mathcal R}$
acting on F such that all the points $(a'_1,\ldots,a'_n)$ are lower than $(a_1,\ldots,a_n)$.

Therefore claiming that the number of master integrals is finite is equivalent to saying that
$\mathcal S$ is a finite-dimensional vector-space.

\section{Reformulation}

We will be using the definition of master integrals from the previous section, however we will need to reformulate it in a more convenient way
to be able to apply some theorems from algebraic geometry.

As it has been stated in \cite{Lee}, the $\mathbb Q$-span of IBPs (before substituting indices) forms a Lie algebra $\mathrm g$. Even more, we can describe this algebra - the $\mathbb Q$-span of
$\partial_{k_i}\cdot k_j$ (where $k_{i,j}$ are loop momenta) forms $\mathfrak{gl}_h(\mathbb Q)$ --- the Lie algebra of $h$ by $h$ matrices. The $\mathbb Q$-span of $\partial_{k_i} \cdot q_j$ (where $q_j$ are external momenta) forms the nilradical of $\mathrm g$, which is commutative and isomorphic, as a $\mathfrak{gl}_h(\mathbb Q)$-module,  to the sum of $N$-copies of the natural representation $\mathbb Q^h$.

% There is a natural representation of $\mathrm g$ on $\mathbb F(q)[k_1,..., k_n]$ --- polynomial functions over $\mathbb F(q)$ of $4\times n$ variables. It is easy to see that $\mathrm g$ acts on the finite-dimensional $\mathbb F(q)$-space $\langle k_ik_j, q_ik_j\rangle_{\mathbb F(q)-\mathrm{span}}$ and therefore on the subring $\mathbb F(q)[k_ik_j, k_iq_j]\subset \mathbb F(q)[k_1,..., k_j]$.

Let $\mathbb C(q)=\mathbb C(q_1,..., q_N)$ be a transcendental extension of $\mathbb C$ by $4\times N$ independent variables. There is a natural action of SO$_{1,3}(\mathbb C)$ on this space and invariants of this action coincide with the subfield $\mathbb C(q_i \cdot q_j)$ where $q_i \cdot q_j$ are the scalar products of $q_i, q_j$. If $N\le 4$ the field $\mathbb C(q_i \cdot q_j)$ is freely generated by $q_i \cdot q_j$. Otherwise this field is freely generated by elements $\{q_mq_n\}_{1\le m,n\le 4}$ and $\{q_bq_m\}_{m\le 4, b>4}$.
Now let us extend this field $\mathbb C(q_i \cdot q_j)$ once more by the dimension $d$ and masses $m_i$.
In what follows we denote $\mathbb C(q_i \cdot q_j,d,m_i)$ by $\mathbb F$.

One more statement which follows from \cite{Lee} is that $\mathrm g(\mathbb F)$ is the tangent algebra to $G(\mathbb F)$ --- the subgroup of $\mathbb F$-linear transformations of the $\mathbb F$-vector space spanned by all $k_i$ and $q_i$ preserving the subspace spanned by $k_i$.
This group acts on Feynman integrals semi-invariantly. The subgroup of elements with a determinant $1$ acts invariantly and the one-dimensional subgroup (the center) multiplying all vectors by $t$ multiplies Feynman integrals by $t^d$.
All elements of $\mathrm g(\mathbb F)$ of the form $\partial_{k_i}\cdot k_j$ where $i\neq j$ and $\partial_{k_i}\cdot q_j$ as well as $\partial_{k_i}\cdot k_i-\partial_{k_j}\cdot k_j$ vanish when applied to Feynman integrals since they are tangent to the subgroup of elements with a determinant $1$.
The sum $\sum_i\partial_{k_i}\cdot k_i$ applied to Feynman integrals multiplies them by $h d$. Let $\chi:\mathfrak g(\mathbb Q)\to \mathbb F$ (where $g\to hd$ tr $g$) be a character. As a result of the statements in this paragraph, we have the following relation\footnote{The
fact that IBPs vanish might be considered as a corollary of two statements --- that $\partial_{k_i}\cdot k_j$ and $\partial_{k_i}\cdot q_j$ form a basis of the tangent algebra of $G$ and that the integrals are semi-invariant under the action of $G$.}:
$$ \forall g\in\mathrm g\hspace{20pt}\mathrm F(a_1,\ldots,a_n) = \int \cdots \int \mbox{d}^d k_1\ldots \mbox{d}^d k_h\mathrm (g-\chi(g))\frac{1}{\mathrm E_1^{a_1}\ldots \mathrm E_n^{a_n}}=0$$

For a set of indices $(a_1,\ldots,a_n)$ there is a function $(\mathrm E_1^{a_1}\ldots \mathrm E_n^{a_n})^{-1}\in\mathbb F(k_1,\cdots, k_h)$. All $\mathrm E_i$ are quadratic forms of variables $k_i$ and $q_j$ (with coefficients  in $\mathbb F$) plus constants; only the products $k_i \cdot k_j$ and $k_i \cdot q_j$ are allowed. Therefore all $\mathrm E_i$ are elements of the ring $\mathbb F[k_i \cdot k_j, k_i \cdot q_j]\subset\mathbb F[k_1,\cdots, k_h]$. The action of the Lie algebra $\mathrm g(\mathbb F)$ on $A:=\mathbb F[k_i \cdot k_j, q_i \cdot k_j]$ extends to the action of the group $G(\mathbb F)$. The action of the group $G(\mathbb F)$ on a finitely generated commutative algebra $A$ over $\mathbb F$ induces the action of $G(\bar{\mathbb F})$ on Spec$_{\bar{\mathbb F}}A=$Hom$_{alg}(A, \bar{\mathbb F})$ where $\bar{\mathbb F}$ is an algebraic closure of $\mathbb F$. Now all $\mathrm E_i$ become functions on Spec$_{\bar{\mathbb F}}A$, therefore $\mathcal F$ can be considered as functions on the complement in Spec$_{\bar{\mathbb F}}A$ to hypersurfaces defined by $\mathrm E_i$. The substitution of indices into IBPs now becomes equivalent to applying $\mathrm g$ to $\mathcal F$. Therefore we should prove that $$(\mathrm g -\chi(\mathrm g))\mathcal F\backslash \mathcal F:=\langle gf-\chi(g)f\rangle_{g\in\mathrm g, f\in\mathcal F}\backslash\mathcal F$$ is a finite-dimensional vector space.

Theorem \ref{Phys} becomes a consequence of the following theorem (the proof is presented in the next section)
\begin{theorem}\label{Math} Let $G$ be an algebraic group acting on a vector space $X$ such that the action $G:X$ has a finite number of orbits. Let $\chi: \mathrm g\to\mathbb F$ be a character (=a homomorphism of Lie algebras). Then for any set of functions $\mathrm E_1,\ldots, \mathrm E_n\in \mathbb F[X]$ the Lie algebra $\mathrm g$ acts on the ring of regular functions $\mathcal F$ on the complement in $X$ to the hypersurfaces defined by $\mathrm E_i$. Then the quotient $(\mathrm g-\chi(\mathrm g)) \mathcal F\backslash \mathcal F$ is finite-dimensional.\end{theorem}

In fact, Theorem \ref{Phys} is not directly sufficient to prove Theorem \ref{Math} for two reasons.
The first of those is that the variety Spec$_{\bar{\mathbb F}}A$ has a finite number of $G(\bar{\mathbb F})$-orbits but not necessary smooth. Nevertheless the statement of Theorem \ref{Math} holds for it because there is a proper inclusion $j:$Spec$_{\bar{\mathbb F}}A\to X$ into a smooth $G(\bar{\mathbb F})$-variety $X$ with a finite number of $G(\bar{\mathbb F})$-orbits.

The second reason is that the algebra of functions generated by $k_i \cdot k_j$ can be not free. Let us prove that our case is a consequence of Theorem \ref{Phys}.

The algebra $A:=\mathbb F[k_i \cdot k_j, k_i \cdot q_j]$ has a $G(\mathbb F)$-stable subalgebra $A_0:=\mathbb F[k_i \cdot q_j]$. Let $\hat A:=A_0[k_i \cdot k_j]$ be a free algebra over $A$ generated by elements $k_i \cdot k_j$. We can define an action of $\mathrm g$ on $\hat A$ using rules $$(1)~(k_{i'}\cdot \partial_{k_{j'}})(k_i \cdot k_j):=\delta_{j', i}(k_{i'}\cdot k_j)+\delta_{j',j}(k_{i'}\cdot k_i),\hspace{20pt}(2)~(q_{i'}\cdot \partial_{k_{j'}})(k_i \cdot k_j):=\delta_{j', i}(q_{i'}\cdot k_j)+\delta_{j',j}(q_{i'}\cdot k_i).$$

The inclusion map $i:A_0\to \hat A$ induces the covering map of $G(\bar{\mathbb F})$-varieties $i$:~Spec$_{\bar{\mathbb F}}\hat A\to$Spec$_{\bar{\mathbb F}}A_0$. The unipotent radical of $G(\bar{\mathbb F})$ controlled by vector fields $\partial_{k_i}\cdot q_j$ acts as the group of parallel transforms of an affine space Spec$_{\bar{\mathbb F}}A_0$. Therefore it acts transitively on Spec$_{\bar{\mathbb F}}A_0$ and then on the set of fibres of $i$. It is enough to prove that some fiber of $i:$Spec$_{\bar{\mathbb F}}\hat A\to$Spec$_{\bar{\mathbb F}}A_0$ intersects only finitely many $G(\bar{\mathbb F})$-orbits.

The fiber $i^{-1}(0)$ coincides with Spec$_{\bar{\mathbb F}}(\hat A/(q_i \cdot k_j)\hat A)$. Hence the Lie algebra $\mathrm{gl}_h(\mathbb F):=k_i\cdot \partial_{k_j}$ preserves the ideal $(q_i \cdot k_j)\hat A$, it acts on the quotient $\hat A/(q_i \cdot k_j)\hat A$ and therefore GL$_h(\bar{\mathbb F})$ acts on Spec$_{\bar{\mathbb F}}(\hat A/(q_i \cdot k_j)\hat A)$. As a GL$_h(\bar{\mathbb F})$-variety Spec$_{\bar{\mathbb F}}(\hat A/(q_i \cdot k_j)\hat A)$ coincide with S$^2(\mathbb F^h)$. Therefore it has a finite number of GL$_h(\bar{\mathbb F})$-orbits (an orbit is a set of bilinear forms with fixed rank). As a corollary the action of $G(\bar{\mathbb F})$ on Spec$_{\bar{\mathbb F}}\hat A$ has a finite number of orbits.

The surjective morphism of algebras $j:\hat A\to A$ induces a proper inclusion of varieties $j:$Spec$_{\bar{\mathbb F}}A\to$Spec$_{\bar{\mathbb F}}\hat A$. If E$_1,\ldots,$E$_n\in A$ is a set of functions then we can choose(any) preimages of them $\hat{\mathrm E}_1,\ldots, \hat{\mathrm E}_n\in\hat A$. Let $\hat{\mathcal F}$ be a set of functions on the complement in Spec$_{\bar{conceptual \mathbb F}}\hat A$ to hypersurfaces defined by $\hat{\mathrm E}_i$ and $\mathcal F$ be a set of functions on the complement in Spec$_{\bar{\mathbb F}}A$ to hypersurfaces defined by E$_i$. Then there is a surjective map $\hat{\mathcal F}\to \mathcal F$ and therefore surjective map $(\mathrm g-\chi(\mathrm g))\hat{\mathcal F}\backslash\hat{\mathcal F}\to(\mathrm g-\chi(\mathrm g))\mathcal F\backslash\mathcal F$. Hence by Theorem~\ref{Math} $(\mathrm g-\chi(\mathrm g))\hat{\mathcal F}\backslash\hat{\mathcal F}$ is a finite-dimensional vector space, the quotient $(\mathrm g-\chi(\mathrm g))\mathcal F\backslash\mathcal F$ is a finite dimensional vector space too.

\section{Proof of Theorem \ref{Math}}

%In this section we are going to prove Theorem \ref{Phys}.
The reader familiar with the technique of holonomic D-modules could consider Theorem~\ref{Math} as an exercise. Below we explain several basic features of this technique and for all details we direct the reader to~\cite{Cou}.

The base field for all objects is an algebraically closed field $\mathbb F$ of characteristic 0. Let $G$ be an algebraic group with a Lie algebra $\mathrm g$, $X$ be a $G$-variety. Let $\mathcal D[X]$ be a ring of polynomial differential operators on $X$. Let $x_1, ..., x_n\in X$ be a basis of $X$ and $\partial_{x_1},  ..., \partial_{x_n}\in X^*$ be a dual basis in vector fields Vec$(X)$. There is a rising (Bernstein) filtration \begin{center}$\bigcup_{i\ge 0}\mathcal D_i:=\langle x_1^{i_1}x_2^{i_2}...x_n^{i_n}\partial_{x_1}^{j_1}...\partial_{x_n}^{j_n}|~i_1+...+i_n+j_1+...+j_n\le i\rangle\subset\mathcal D[X]$\hspace{40pt}.\end{center}One can easily see that the graded algebra $\oplus_{i\ge 0} \mathcal D_{i}/\mathcal D_{i-1}$ with respect to this filtration is isomorphic to $\mathbb F[$T$^*X$] and $\mathcal D_i$ is a finite-dimensional vector space for all $i$.

Let $\mathcal F$ be a finitely generated $\mathcal D[X]$-module with a finite-dimensional generating space $\mathcal F_0$. Let $\mathcal F_i:=\mathcal D_i \mathcal F_0$. The graded space gr$\mathcal F:=\oplus_{i\ge 0}\mathcal F_{i+1}/\mathcal F_i$ is a module over $\mathbb F[$T$^*X]$ and could be considered as a sheaf over T$^*X/\mathbb F$. The support of this sheaf V$(\mathcal F)\subset$T$^*X\cong X\oplus X^*$ does not depend on $\mathcal F_0$.
\begin{theorem}[J. Bernstein,{\upshape~\cite{Cou}[\S~9.4]}]Let $\mathcal F$ be a finitely generated $\mathcal D[X]$-module. Then the dimensions of all irreducible components of the variety $\mathrm V(\mathcal F)$ are not less than $n$.\end{theorem}
\begin{definition}A finitely generated $\mathcal D[X]$-module $\mathcal F$ is called holonomic if the dimensions of all irreducible components of the variety $\mathrm V(\mathcal F)$ are equal to $n$.\end{definition}
Let $\mathcal F_1, \mathcal F_2$ be a $\mathcal D[X]$-modules. Then the module $$\mathcal F_1\otimes_{\mathbb F[X]} \mathcal F_2:=\mathcal F_1\otimes_{\mathbb F} \mathcal F_2/\langle f\cdot f_1\otimes f_2-f_1\otimes f\cdot f_2\rangle_{f_1\in\mathcal F_1, f_2\in\mathcal F, f\in\mathbb F[X]}$$ is a $\mathcal D[X]$-module where an action of vector fields is defined by the Leibniz rule $\xi (f_1\otimes f_2)=\xi f_1\otimes f_2+f_1\otimes \xi f_2$.
\begin{lemma}[{\upshape~\cite{Cou}[Ex.~4, \S 16.4]}]\label{Fpi} Suppose that $\mathcal F_1, \mathcal F_2$ are finitely generated holonomic $\mathcal D[X]$-modules. Then $\mathcal F_1\otimes_{\mathbb F[X]}\mathcal F_2$ is a finitely generated holonomic $\mathcal D[X]$-module.\end{lemma}
Let $\mathcal F$ be a $\mathcal D[X]$ module. The quotient $\pi^+_{\mathrm{pt}}\mathcal F=\mathcal F/({\partial_{x_1}\mathcal F+...+\partial_{x_n}\mathcal F})$ is a vector space (=$\mathcal D$-module over a point).
\begin{lemma}[{\upshape~\cite{Cou}[$\S$~16.3]}]\label{pi} Suppose $\mathcal F$ is a finitely generated holonomic $\mathcal D[X]$-module. Then $\pi^+_{\mathrm{pt}}\mathcal F$ is a finite-dimensional vector space (=finitely generated holonomic $\mathcal D$-module over the point).\end{lemma}
There is an involutive antiautomorphism $\phi:\mathcal D[X]\to\mathcal D[X]$ such that $\forall \alpha, \beta ~\phi(x^\alpha\partial_x^\beta)=(-1)^{|\alpha|}x^\beta\partial_x^\alpha$ where $\alpha, \beta$ are multi-indexes. Involutive antiautomorphism turn left-modules to right-modules and vice versa.
\begin{lemma}~\label{product} Let $\mathcal F_1$ be a right and $\mathcal F_2$ be a left finitely generated holonomic $\mathcal D[X]$-modules. Then $$\mathcal F_1\otimes_{\mathcal D[X]}\mathcal F_2:=F_1\otimes_{\mathbb F} F_2/\langle f_1\cdot f\otimes f_2-f_1\otimes f\cdot f_2\rangle_{f_1\in\mathcal F_1, f_2\in\mathcal F, f\in\mathcal D[X]}$$ is a finite dimensional vector space.\end{lemma}
\begin{proof}There are two surjective maps $\alpha_1: \mathcal F_1\otimes_\mathbb F\mathcal F_2\to \mathcal F_1\otimes_{\mathcal D[X]}\mathcal F_2$ and $\alpha_2: \mathcal F_1\otimes_\mathbb F\mathcal F_2\to\pi^+_{\mathrm{pt}}(\mathcal F_1^\phi\otimes_{\mathbb F[X]}\mathcal F_2)$. They have the same kernel. The image of $\alpha_2$ is finite-dimensional by lemmas~\ref{Fpi} and~\ref{pi}. Then the image of $\alpha_1$ is finite-dimensional too. \end{proof}
Let $\mathrm g$ be a Lie algebra and $W$ be a $\mathrm g$-module (not necessary finite-dimensional!).
\begin{definition}The action of $\mathrm g$ on $\mathcal F$ is called locally finite if $\forall m\in \mathcal F$ there exists a finite-dimensional $\mathrm g$-submodule $V\subset \mathcal F$ such that $~m\in V$.\end{definition}
Let $K$ be an algebraic group over $\mathbb F$ with a Lie algebra $\mathrm g$. Let $X$ be an $\mathbb F$-vector space with an algebraic action of $K$ and $\mathcal F$ be a finitely generated $\mathcal D[X]$-module.
\begin{lemma}Suppose that the action of $\mathrm g$ on $\mathcal F$ is locally finite. Then $\mathrm V(\mathcal F)$ is annihilated by $\mathrm g\subset (X\oplus\mathbb C)\otimes X^*\subset\mathbb F[\mathrm T^*X]\cong \mathbb F[X\oplus X^*]$. In particular, $\mathrm V(\mathcal F)$ lies in the union of conormal bundles to all $\mathrm g$-orbits $$\mathrm N_{\mathrm g, X}:=\{(x, l)\in X\oplus X^*| l(\mathrm gx)=0\}$$.\end{lemma}
We are ready to prove Theorem~\ref{Math}:
\begin{proof}[Proof of Theorem~\ref{Math}]Obviously $(\mathrm g-\chi(\mathrm g)) \mathcal F\backslash \mathcal F=(\mathrm g-\chi(\mathrm g)) \mathcal D[X]\backslash\mathcal D[X]\otimes_{\mathcal D[X]} \mathcal F$. The Lie algebra $\mathrm g$ acts on $$(\mathrm g-\chi(\mathrm g)) \mathcal D[X]\backslash\mathcal D[X]:=\langle gd-\chi(g)d\rangle_{g\in\mathrm g, d\in \mathcal D[X]}\backslash\mathcal D[X]$$ locally finitely  and therefore V$(\mathcal F)$ is a subset of N$_{\mathrm g, X}\subset$T$^*X$. Hence there are only finitely many $\mathrm g$-orbits on $X$, the variety N$_{\mathrm g, X}$ is equidimensional of the dimension dim~$X$. Therefore the module $(\mathrm g-\chi (\mathfrak g)) \mathcal D[X]\backslash\mathcal D[X]$  is  holonomic. Since $\mathcal F$ is a finitely generated holonomic $\mathcal D[X]$-module~\cite{Cou}[Ex 2, \S 18.4] $(\mathrm g-\chi(\mathrm g)) \mathcal D[X]\backslash\mathcal D[X]\otimes_{\mathcal D[X]} \mathcal F$ is a finite-dimensional vector space by Lemma~\ref{product}.\end{proof}

\section{Conclusion}

In this paper we show that the number of master integrals is always
finite. We should note that our proof is valid for standard Feynman
integrals but for some `exotic' integrals invovling infinite summations (see e.g. \cite{ftft}),
the statement does not hold, see an example of integrals in
finite-temperature field theory in \cite{Schroder:2008ex}.

Our proof is non-constructive, meaning that we derive no direct
restrictions on the number of master integrals or there possible values.
However we plan to continue our research in order to obtain practical
hints for Feynman integral reduction.


\begin{thebibliography}{99}
\bibitem{dimreg}
G.~'t Hooft and M.~Veltman,
Nucl.~Phys. {\bf B 44} (1972) 189;
%%CITATION = NUPHA,B44,189;%%
\\
C.~G.~Bollini and J.~J.~Giambiagi,
Nuovo Cim. {\bf 12 B} (1972) 20.
%%CITATION = NUCIA,B12,20;%%

%\bibitem{Smirnov:2004ym}
%  V.~A.~Smirnov,
%  ``Evaluating Feynman Integrals,''
%  Springer Tracts Mod.\ Phys.\  {\bf 211} (2004) 1ю
%%CITATION = STPHB,211,1;%%

\bibitem{Smirnov:2006ry}
V.~A.~Smirnov,
  ``Feynman integral calculus,''
{\it  Berlin, Germany: Springer (2006) 283 p}.

\bibitem{IBP}
K.~G.~Chetyrkin and F.~V.~Tkachov,
Nucl. Phys. {\bf B 192}  (1981) 159.

\bibitem{LGR1}
S.~Laporta,
%{\em High-precision calculation of multi-loop Feynman
%integrals by difference equations,}
Int. J. Mod. Phys. {\bf A 15} (2000) 5087 [hep-ph/0102033].

\bibitem{LGR2}
T.~Gehrmann and E.~Remiddi,
%{\em Two-loop master integrals for $\gamma^*$ $\to$ 3 jets:
%the planar  topologies,}
Nucl. Phys. {\bf B 601} (2001) 248 [hep-ph/0008287];
\\
T.~Gehrmann and E.~Remiddi,
%{\em Two-loop master integrals for $\gamma^*$ $\to$ 3 jets:
%the non-planar  topologies,}
Nucl. Phys. {\bf B 601} (2001) 287 [hep-ph/0101124].


\bibitem{AIR}
C.~Anastasiou and A.~Lazopoulos,
%{\em Automatic integral reduction for higher order perturbative calculations}
JHEP {\bf 0407} (2004) 046 [hep-ph/0404258].

\bibitem{FIRE}
A.~V.~Smirnov,
%{\em Algorithm FIRE - Feynman Integral Reduction} //
JHEP {\bf 0810} (2008) 107 [0807.3243]

\bibitem{Reduze}
C.~Studerus, [0912.2546]

\bibitem{masters} A.~V.~Smirnov and V.~A.~Smirnov,
%On the reduction of Feynman integrals to master integrals,
PoS ACAT 2007:085 [0707.3993]

\bibitem{Lee} R.~N.~Lee,
%Group structure of the integration-by-part identities and its application to the reduction of multiloop integrals,
JHEP {\bf 0807} (2008) 031 [0804.3008]

\bibitem{Cou} Coutinho, S.C. ``A primer of algebraic D-modules'', London Math. Soc. student texts {\bfseries 33}, Cambridge University Press, 1995

\bibitem{ftft}
 J.~I.~Kapusta and C.~Gale,
 ``Finite-Temperature Field Theory Principles and Applications,''
 Cambridge University Press, second edition, 2006.

\bibitem{Schroder:2008ex}
  Y.~Schr\"oder,
  %``Loops for Hot QCD,''
  Nucl.\ Phys.\ Proc.\ Suppl.\  B {\bf 183} (2008) 296
  [arXiv:0807.0500 [hep-ph]].
  %%CITATION = NUPHZ,B183,296;%%


\end{thebibliography}
\end{document}